\documentclass{article}

\begin{document}

\title{Null energy in plane Kaluza-Klein worlds}
\author{David Solano$^{1,2\,\,a}$ and Rodrigo Alvarado$^{1\,\,b}$ \thanks{%
electronic address: \newline
$^{a}$dsolano@efis.ucr.ac.cr, dsscr@yahoo.com; \newline
$^{b}$realmar@cariari.ucr.ac.cr} \\
$^{1}$Escuela de Fisica, Universidad de Costa Rica\\
Cuidad Universitaria Rodrigo Facio, Montes de Oca, San Jose.\\
and\\
$^{2}$Laboratorio Nacional de Nanotecnologia, \ \\
Centro Nacional de Alta Tecnologia. \\
1 km south from the U.S. Embassy \\
Pavas. San Jose, Costa Rica}
\maketitle

\begin{abstract}
We study the energy problem in the $R0X$ spacetimes in 5D. We found that for
the Einstein, Landau-Lifschitz and the Moller complexes are null.
\end{abstract}

\newpage

\section{Introduction}

Energy is standard definition that comes from the born of the analytical
mechanics that makes easier the study of the dynamics of a physical system.
We can always (with few exceptions) define certain conserved quantity
--conjugated to coordinate time-- that allow us to perform predictions and
establish relation with other dynamical measurable observables. Although, in
the very foundations of the General Relativity the definition of energy is
rather than difficult sometimes uncertain. As it is well-known, the general
relativist abandons the concept of force by using the principle of free-fall
observers. In a mathematical language, we can write a 4D version of the
Newton law of motion in curved spacetime where the ``acceleration'' is null

\begin{equation}
\frac{D^{2}a^{\mu }}{D\tau ^{2}}=0.  \label{newton}
\end{equation}

Therefore, we change our dynamical vision of the interaction as force by
curvature:

\begin{equation}
\frac{D^{2}n^{\mu }}{D\tau ^{2}}+R_{\nu \rho \tau }^{\mu }\frac{d\xi ^{\nu }%
}{d\tau }n^{\rho }\frac{d\xi ^{\tau }}{d\tau }=0,  \label{geo_des}
\end{equation}

where $n^{\mu }$ is relative position of two near particle trayectories
(geodesics), $\xi ^{\nu }$ the coordinate in the manifold and $R_{\nu \rho
\tau }^{\mu }$ is the Riemann curvature tensor. Since no longer the concept
of force is compatible with the General Relativity, we cannot perform
addecuately a definition of energy. For instance, this problem have made
difficult the journey for those who attempt to construct a consistent and
complete quantum theory of gravitation \cite{dewitt}, mainly because every
quantum theory needs a well-defined Hamiltonian operator and the Hamiltonian
constraint for GR is null for every dynamical solution.

This problem occupied, first, Einstein himself \cite{einstein}, who proposed
a complex of gravitational Energy-Momentum which main technical disadvantage
is that it is not tensor by definition. The Einstein complex do not satisfy
a fully covariant local conservation property (a null divergence), so as
calculation tool is efficient only in Cartesian coordinates, as observed by
Xulu \cite{xulu}, who has exhaustive study this topic. In order to solve
some Einstein complex, Landau and Lifshitz \cite{landau} propose a complex
with tensor properties only on geodesic coordinate systems (where all $%
\partial g_{\mu \nu }/\partial \xi ^{\rho }=0$). Finally, and not less,
Moller \cite{moller} derived a complex that generates a energy-momentum
4-vector that transform explicitly as that. Others complexes of
gravitational energy has been created in the literature in order to give a
more complete analysis of the interaction, like Papapetrou-Gupta \cite{papa}%
--who uses the spin operator of the Einstein-Hilbert Lagrangian but mixes
the flat Minkowski geometry-- and Weinberg \cite{weinberg}-- specially for
weak field approximation--.

In \cite{david}, we call \textit{R0X} spacetimes in $N$ dimensions to those
vacuum solutions to Einstein equations that \textit{1.-} posses plane
symmetry, \textit{2.-} own a particular structure that depends only on a
arbitrary function we named generating function and a set of parameters that
define a manifold with $N-3$ dimension, and \textit{3.-} have exact
solutions to the geodesic equations. The ``X'' can be ``T'' when the
generating function is well-tempered, ``B'' when it is bounded, ``U'' when
it is unbounded and ``L'' when it has a logaritmic structure. In this paper,
we show some interesting results about the behavior of energy-momentum\ in
R0X spacetimes (with $N=5$) by making use of the Einstein, Landau and Moller
complexes. The R0X metric we use has the general form

\begin{equation}
ds^{2}=e^{A_{0}\chi }dt^{2}-\left( e^{A_{1}\chi }dx^{2}+e^{A_{2}\chi
}dy^{2}+e^{A_{3}\chi }dz^{2}\right) -\left( \frac{d\chi }{d\eta }\right)
^{2}e^{\chi }d\eta ^{2}  \label{metric}
\end{equation}

with the usual convention $\xi ^{0}=t,\xi ^{1}=x,\xi ^{2}=y,\xi ^{3}=z,\xi
^{4}=\eta $.

\section{Results and conclusions}

For both Einstein and Moller energy complexes, we need to define the
non-tensorial quantity:

\begin{equation}
h_{B}^{\quad CD}=\frac{g_{BM}}{\sqrt{g}}\frac{\partial }{\partial \xi ^{N}}%
\left[ g\left( g^{CM}g^{DN}-g^{DM}g^{CN}\right) \right]  \label{h_einstein}
\end{equation}

The Einstein energy momentum complex is then given by (in natural units):

\begin{equation}
\Theta _{B}^{C}=\frac{1}{16\pi }\frac{\partial }{\partial \xi ^{D}}%
h_{B}^{\quad CD}  \label{einstein_en}
\end{equation}

and the Moller complex is

\begin{equation}
S_{B}^{C}=\frac{1}{16\pi }\frac{\partial }{\partial \xi ^{D}}\left( \delta
_{B}^{C}h_{M}^{\quad MD}-\delta _{B}^{D}h_{M}^{\quad MC}-2h_{B}^{\quad
CD}\right)  \label{moller_en}
\end{equation}

For the metric in (\ref{metric}), the non-null Einstein pseudo tensor
components are (for $C=0,1,2,3$)

\begin{equation}
\Theta _{C}^{C}=\frac{1}{32\pi }\frac{d\chi }{d\eta }\left(
1-\sum_{B=0}^{4}A_{B}\right) \left( \sum_{B=0,B\neq C}^{4}A_{B}\right)
\label{r1}
\end{equation}

and for the Moller one:

\begin{equation}
S_{C}^{C}=\frac{1}{16\pi }\frac{d\chi }{d\eta }\left(
1-\sum_{B=0}^{4}A_{B}\right) \left( 1+2A_{C}\right)  \label{r2}
\end{equation}

From \cite{david} and \cite{david2}, the $A_{B}$ parameters satisfy the
relation $\sum_{B=0}^{4}A_{B}=1$, a direct consequence of the Einstein field
equations. Therefore, for every R0X the Einstein and Moller complexes are
null.

Now, for the Landau-Lifshitz energy momentum complex given in \cite{landau}
and \cite{xulu}, we have after some simplifications

\begin{equation}
L_{C}^{C}=\frac{1}{32\pi }e^{\chi }\left( \frac{d\chi }{d\eta }\right)
^{2}\left( A_{C}-1\right) \left[ 2\frac{d^{2}\chi }{d\eta ^{2}}\left( \frac{%
d\chi }{d\eta }\right) ^{-1}-A_{C}-\frac{1}{2}\right]  \label{r3}
\end{equation}

Landau-Lifshitz energy-momentum definition is valid only for geodesic
coordinate system ($\partial g_{BC}/\partial \xi ^{D}=0$). Therefore, a
geodesic coordinate system is that satisfies the following two conditions
from the derivatives of the metric tensor (\ref{metric}) at certain $\eta
_{0}$

\begin{eqnarray}
\left. \frac{d\chi }{d\eta }\right| _{\eta _{0}} &=&0  \label{condi1} \\
2\left( \left. \frac{d^{2}\chi }{d\eta ^{2}}\right| _{\eta _{0}}\right)
+\left( \left. \frac{d\chi }{d\eta }\right| _{\eta _{0}}\right) ^{2} &=&0
\label{condi2}
\end{eqnarray}

Notice that these conditions ensure $\left. d\chi /d\eta \right| _{\eta
_{0}}=\left. d^{2}\chi /d\eta ^{2}\right| _{\eta _{0}}=0$. It is easy to
picture spacetimes that could contain one or more geodesic coordinate system
from the freedom to choose the generating function. For instance, a periodic
generating function \cite{david} that produces a R0B spacetime would have a
infinite number of geodesic coordinate systems. Thus, $L_{C}^{C}=0$, in any
geodesic coordinate system.

If we extend our analysis to R0L spaces where the generating function has a
the generic form $\chi =\ln \theta $, where $\theta $ is a function of $\eta 
$, some new subtleties come out from the functional structure of the metric.
The conditions for geodesic systems becomes then

\begin{eqnarray}
\left[ \theta \left( \eta _{0}\right) \right] ^{A_{0}-1}\left. \frac{d\theta 
}{d\eta }\right| _{\eta _{0}} &=&0  \label{condi3} \\
\left. \frac{d\theta }{d\eta }\right| _{\eta _{0}}\left[ 2\theta \left( \eta
_{0}\right) \left( \left. \frac{d^{2}\theta }{d\eta ^{2}}\right| _{\eta
_{0}}\right) -\left( \left. \frac{d\theta }{d\eta }\right| _{\eta
_{0}}\right) ^{2}\right]  &=&0  \label{condi4}
\end{eqnarray}

To avoid problems with infinities at the Christoffel connections, and thus
with the definition of energy, let be $\theta \left( \eta _{0}\right) \neq 0$
for any suitable geodesic system. Therefore, $\left. d\theta /d\eta \right|
_{\eta _{0}}=0$ is a sufficient condition that a geodesic coordinate system
satisfy and that $\left. d^{2}\theta /d\eta ^{2}\right| _{\eta _{0}}$ is not
necessary to be $0$. For R0L spaces the conditions for constructing geodesic
coordinate systems differ from the others by a sign. Then, the
Landau-Lifshitz complex for R0L spacetimes is

\begin{eqnarray}
L_{C}^{C} &=&\frac{1}{16\pi \theta }\left( A_{C}-1\right) \left[ \theta 
\frac{d^{2}\theta }{d\eta ^{2}}-A_{C}\left( \frac{d\theta }{d\eta }\right)
^{2}\right]   \label{r4} \\
&=&\frac{1}{16\pi }\left( A_{C}-1\right) \frac{d^{2}\theta }{d\eta ^{2}} 
\end{eqnarray}

So, in R0L spacetimes the Landau-Lifshitz complex is not necessary null
unless we find a $\theta $-function that certain $\eta _{0}$ is both an
extreme and an inflexion point. In other hand, the gravitational energy
density in the ($L_{t}^{t}$) is null always if the spacetime structure
constant $C_{0}=1$, \ a possible value for the specific parametrization of $%
N=5$ R0X as we can see in \cite{david2}.

Let summarize our results. In principle, the gravitational complexes are
artificial structures that emulate a energy-momentum tensor $T_{BC}$ for the
gravitational interaction. Thus, in Kaluza-Klein theories we must find the
most appropriate interpretation to each its components over the
extra-dimensions. By following \cite{wesson} and \cite{no_tl}, a time-like
signature for the extra-dimension $\eta $ would imply several undesirable
features, like the wrong sign in the 4D Maxwell action relative to the
Einstein one, the prediction of tachyons and closed timel-ike curves that
allow causality violations. Hence, the choice of space-like signature for $%
\eta $ ensures no inconsistencies with our standard ideas about spacetime.

Thus, by analogy with the standard definition of the energy-momentum tensor
in the standard four-dimensional spacetime, $T_{\eta i}$ ($i=1,2,3$) are the
proyection on the $\eta $-direction of forces that act on matter across a
unit surface with normal vector $\mathbf{e}_{i}$. In other words, $T_{\eta
i} $ is the $\eta $-components of the momentum flux in the direction $i$.
Thus, our result ($T_{i}^{\eta }=0$) is consistent with the intuitive idea
that there is no matter momentum flux over the extra-dimension. In the order
side, $T_{t\eta }$ is energy flux on the $\eta $-direction and as we should
expect, there is no energy interchange into the ``hidden'' dimension.

Since all energy-momentum components in all the three complexes we have
studied are null, we can conjecture that when we unify the theory of
gravitation and the electromagnetism the ``effective'' energy of hole system
is null for any R0X spacetime in 5D. Hence, for a self-gravitating
electromagnetic system described by the Kaluza-Klein 5D vacuum gravity,
there could exist a balance between the two interactions that prevents the
formation of gravitational effective masses, in a similar matter that do
Bianchi I spacetimes \cite{bianchi}. Therefore, quantities like linear and
angular momentum are always conserved with trivial values. We must say that
in the context of non-compactified Kaluza-Klein theories (where the
extra-dimension is not compact, but intuitively must small) other R0X
solutions can be admitted, for instance, all those which generating function 
$\chi =\chi \left( \xi ^{\nu }\right) $ ($\nu =0,1,2,3$). These R0X
solutions that depends on conventional four-dimensional parameters still
conserve the property of null gravitational energy, as one can easily verify
by the extreme symmetry of the plane coordinates we choose for the metric.

Then, R0X worlds are energetically equivalent to vacuum universes.

Acknowledgment. (DS) would like to thank Jorge A. Diaz and Ms. Eda Maria
Arce for their kindness and hospitality at LANOTECH.


\begin{thebibliography}{99}
\bibitem{dewitt}  B. S. DeWitt. Phys Rev \textbf{160} (1967) 1113.

\bibitem{einstein}  A. Einstein. Ann Phys \textbf{49} (1916). For an english
version: \textit{The Foundation of the General Theory of Relativity}, in: H.
A. Lorentz, A. Einstein, H. Minkowski and H. Weyl. \textit{The Principle of
Relativity}. Dover. New York. 1952.

\bibitem{xulu}  S. S. Xulu. The Energy-Momentum Problem in General
Relativity. Ph.D Thesis. University of Zululand. 2002. Online version: 
\texttt{hep-th/0308070} and references therein.

\bibitem{landau}  L. D. Landau and E. M. Lifshitz. \textit{The Classical
Theory of Fields}. Pergamon. London. 1959.

\bibitem{moller}  C. Moller. Ann Phys (NY) \textbf{4} (1958) 347; Ann Phys
(NY) \textbf{12} (1961) 118.

\bibitem{papa}  A. Papapetrou. Proc Roy Ir Ac A \textbf{52 }(1948) 11. S N
Gupta. Phys Rev \textbf{96} (1954) 1683.

\bibitem{weinberg}  S. Weinberg. \textit{Gravitation and Cosmology:
Principles and Applications of General Theory of Relativity}. John Wiley and
Sons. New York. 1972.

\bibitem{david}  D. Solano. \textit{Plane curved spacetimes in }$N$\textit{\
dimensions: General Properties and Definitions}, (in preparation).

\bibitem{david2}  D. Solano and R. Alvarado. \textit{Plane 5D world and
simple compactification}. Online preprint: \texttt{gr-qc/0505127.}

\bibitem{wesson}  J. M. Overduim and P. S. Wesson. Phys Rep \textbf{283}
(1997) 303, online version: \texttt{gr-qc/9805018 }.

\bibitem{no_tl}  D. Bailin and A. Love. Rep Prog Phys \textbf{50} (1987)
1087.

\bibitem{bianchi}  S. S. Xulu. Int J Mod Phys \textbf{A15} (2000) 4849 .
\end{thebibliography}
\end{document}